\documentclass[sttt]{svjour}
\usepackage{graphicx}
\usepackage{enumitem}
\usepackage{color}
\usepackage{url}
\usepackage{longtable}
\usepackage{balance}
\begin{document}
\title{A taxonomy of risk-based testing}

\author{Michael Felderer\inst{1}, Ina Schieferdecker\inst{2}}
\date{}

\institute{
University of Innsbruck, Innsbruck, Austria \\ \email{michael.felderer@uibk.ac.at} \and
Fraunhofer Institute FOKUS and Freie Universit{\"a}t Berlin, Germany \\ \email{ina.schieferdecker@fokus.fraunhofer.de}
}

%
%
\maketitle

\begin{abstract}
Software testing has often to be done under severe pressure due to limited resources and a challenging time schedule facing the demand to assure the fulfillment of the software requirements. In addition, testing should  unveil those software defects that harm the mission-critical functions of the software. Risk-based testing uses risk (re-)assessments to steer all phases of the test process in order to optimize testing efforts and limit risks of the software-based system. Due to its importance and high practical relevance several risk-based testing approaches were proposed in academia and industry. This paper presents a taxonomy of risk-based testing providing a framework to understand, categorize, assess, and compare risk-based testing approaches to support their selection and tailoring for specific purposes. The taxonomy is aligned with the consideration of risks in all phases of the test process and consists of the top-level classes risk drivers, risk assessment, and risk-based test process. The taxonomy of risk-based testing has been developed by analyzing the work presented in available publications on risk-based testing. Afterwards, it has been applied to the work on risk-based testing presented in this special section of the International Journal on Software Tools for Technology Transfer.
\end{abstract}


\section{Introduction} \label{sec:intro}

Testing of safety-critical, security-critical or business-critical software faces the problem of determining those tests that assure the essential properties of the software and have the ability to unveil those software defects that harm the mission-critical functions of the software. However, also for normal, less critical software a comparable problem exists: Usually testing has to be done under severe pressure due to limited resources and tight time constraints with the consequence that testing efforts have to be focused. 

Both decision problems can adequately be addressed by risk-based testing approaches which consider risks of the software product as the guiding factor to steer all phases of the test process, i.e., test planning, design, implementation, execution, and evaluation~\cite{gerrard2002risk,schieferdecker2012MBST,felderer2014integrating}. Risk-based testing is a pragmatic, in industry widely used approach to address the problem of tests for mission-critical software or to cope with ever limited testing resources. Risk-based testing uses the straightforward idea to focus test activities on those scenarios that trigger the most critical situations for a software system~\cite{wendland2012systematic}. 

Due to its importance and high practical relevance several risk-based testing approaches were proposed in academia, e.g.,~\cite{chen2002specification,stallbaum2007employing,zimmermann2009risk,bai2012risk,yoon2011test,wendland2012systematic,felderer2012integrating,zech2012towards,felderer2014integrating}, and industry, e.g.,~\cite{bach1999heuristic,rosenberg2000risk,Amland2000RBT,gerrard2002risk,redmill2004exploring,veenendaal2012prisma}. Recently, the international standard ISO/IEC/IEEE 29119 Software Testing~\cite{ISO2013SoftwareTesting} on testing techniques, processes and documentation even explicitly considers risks as an integral part of the test planning process. Because of the growing number of available risk-based testing approaches and its increasing dissemination in industrial test processes~\cite{felderer2014framework}, support to categorize, assess, compare, and select risk-based testing approaches is required. 

In this paper, we present a taxonomy of risk-based testing providing a framework to understand, categorize, assess, and compare risk-based testing approaches to support their selection and tailoring for specific purposes. In general, a \emph{taxonomy} or classification (scheme) defines a hierarchy of classes (also referred to as categories, dimensions, criteria or characteristics) to categorize things or concepts. It describes a tree structure whose leaves define concrete values to characterize instances. The proposed taxonomy is aligned with the consideration of risks in all phases of the test process and consists of the top-level classes \emph{risk drivers} (with subclasses functionality, safety, and security), \emph{risk assessment} (with subclasses risk item type, factors, estimation, and degree of automation), and \emph{risk-based test process} (with subclasses risk-based test planning, design, implementation, execution, and evaluation). The taxonomy of risk-based testing has been developed by analyzing the work presented in ~\cite{bach1999heuristic,rosenberg2000risk,Amland2000RBT,chen2002specification,redmill2004exploring,redmill2005theory,stallbaum2007employing,stallbaum2008automated,souza2009measurement,zimmermann2009risk,hosseingholizadeh2010source,souza2010risk,kloos2011risk,yoon2011test,zech2011risk,bai2012risk,felderer2012integrating,veenendaal2012prisma,wendland2012systematic,zech2012towards,felderer2013experiences,ray2013risk,felderer2014integrating}. Afterwards, it has been applied to the work on risk-based testing~\cite{Neubauer2014ActiveContinuousQualityControl,Carrozza2014DynamicTestPlanning,Felderer2014MultipleCaseStudy,Erdogan2014SLR} presented in this special section of the International Journal on Software Tools for Technology Transfer (STTT, see Section~\ref{sec:sttt}).

The remainder of this paper is structured as follows. Section~\ref{sec:concepts} presents basic concepts of risk-based testing. Section~\ref{sec:taxonomy} introduces the taxonomy of risk-based testing. Section~\ref{sec:sttt} presents the articles of the STTT Special Section on Risk-Based Testing and discusses them in the context of the taxonomy. Finally, Section~\ref{sec:conclusion} concludes this paper.

\section{Basic concepts of risk-based testing} \label{sec:concepts}

\emph{Testing} is the evaluation of software by observing its execution~\cite{Ammann2008IntroductionToSoftwareTesting}. The executed software-based system is called \emph{system under test} (SUT). \emph{Risk-based testing} (RBT) is a testing approach which considers risks of the software product as the guiding factor to support decisions in all phases of the test process~\cite{gerrard2002risk,schieferdecker2012MBST,felderer2014integrating}. A \emph{risk} is a factor that could result in future negative consequences and is usually expressed by its likelihood and impact~\cite{istqb2012standardGlossary}. In software testing, the \emph{likelihood} is typically determined by the probability that a failure assigned to a risk occurs, and the \emph{impact} is determined by the cost or severity of a failure if it occurs in operation. The resulting \emph{risk value} or \emph{risk exposure} is assigned to a \emph{risk item}. In the context of testing, a risk item is anything of value (i.e., an asset) under test, for instance, a requirement, a component or a fault. 

RBT is a testing-based approach to risk management which can only deliver its full potential if a test process is in place and if risk assessment is integrated appropriately into it. A \emph{test process} comprises the core activities test planning, test design, test implementation, test execution, and test evaluation~\cite{istqb2012standardGlossary}. In the following, we explain the particular activities and associated concepts in more detail.

According to~\cite{IEEE829} and~\cite{istqb2012standardGlossary}, \emph{Test planning} is the activity of establishing or updating a test plan. A test plan is a document describing the scope, approach, resources, and schedule of intended test activities. It identifies, amongst others, objectives, the features to be tested, the test design techniques, and exit criteria to be used and the rationale of their choice. \emph{Test objectives} are reason or purpose for designing and executing a test. The reason is either to check the functional behavior of the system or its nonfunctional properties. \emph{Functional testing} is concerned with assessing the functional behavior of an SUT, whereas \emph{nonfunctional testing} aims at assessing nonfunctional requirements such as security, safety, reliability or performance. The scope of the features to be tested can be components, integration or system. At the scope of \emph{component testing} (also referred to as unit testing), the smallest testable component, e.g., a class, is tested in isolation. \emph{Integration testing} combines components with each other and tests those as a subsystem, that is, not yet a complete system. In \emph{system testing}, the complete system, including all subsystems, is tested. \emph{Regression testing} is the selective retesting of a system or its components to verify that modifications have not caused unintended effects and that the system or the components still comply with the specified requirements~\cite{Radatz1990IEEEStandardGlossary}. \emph{Exit criteria} are conditions for permitting a process to be officially completed. They are used to report against and to plan when to stop testing. Coverage criteria aligned with the tested feature types and the applied test design techniques are typical exit criteria. Once the test plan has been established, test control begins. It is an ongoing activity in which the actual progress is compared against the plan which often results in concrete measures.  

During the \emph{test design} phase the general testing objectives defined in the test plan are transformed into tangible test conditions and abstract test cases. 
\emph{Test implementation} comprises tasks to make the abstract test cases executable. This includes tasks like preparing test harnesses and test data, providing logging support or writing test scripts which are necessary to enable the automated execution of test cases. In the \emph{test execution} phase, the test cases are then executed and all relevant details of the execution are logged and monitored. Finally, in the \emph{test evaluation} phase the exit criteria are evaluated and the logged test results are summarized in a test report.

\emph{Risk management} comprises the core activities \emph{risk identification}, \emph{risk analysis}, \emph{risk treatment}, and \emph{risk monitoring}~\cite{ASNZS2004RiskManagement}. In the risk identification phase, risk items are identified. In the risk analysis phase, the likelihood and impact of risk items and, hence, the risk exposure is estimated. Based on the risk exposure values, the risk items may be prioritized and assigned to risk levels defining a risk classification. In the risk treatment phase the actions for obtaining a satisfactory situation are determined and implemented. In the risk monitoring phase the risks are tracked over time and their status is reported. In addition, the effect of the implemented actions is determined. The activities risk identification and risk analysis are often collectively referred to as \emph{risk assessment}, while the activities risk treatment and risk monitoring are referred to as \emph{risk control}.

\section{Taxonomy of risk-based testing} \label{sec:taxonomy}

\begin{figure*}
    \includegraphics[width=.73\textwidth]{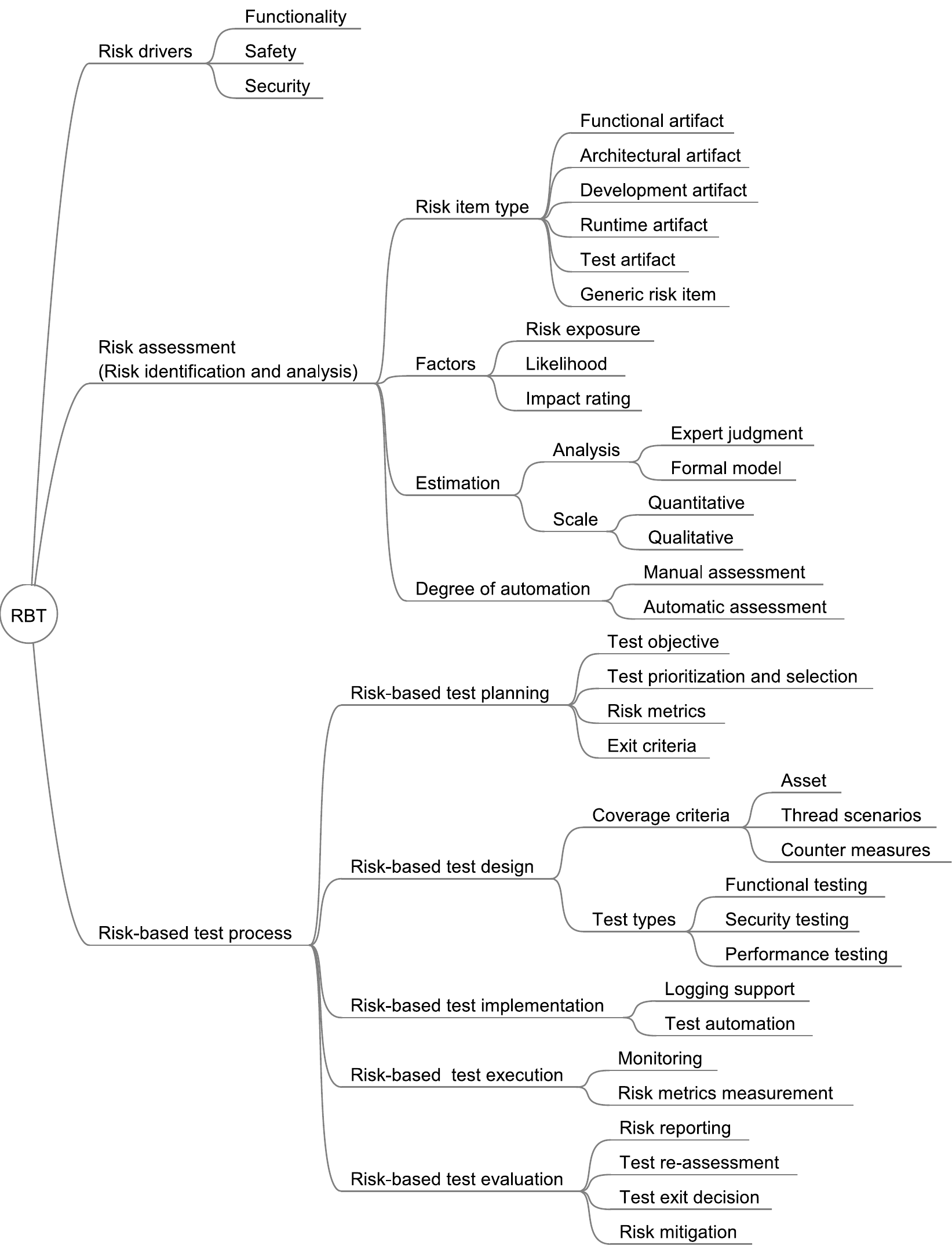}
    \caption{Risk-based testing taxonomy}
    \label{fig:rbt-taxonomy}
\end{figure*}

The taxonomy of risk-based testing is shown in Fig.~\ref{fig:rbt-taxonomy}. It contains the top-level classes \emph{risk drivers}, \emph{risk assessment} as well as \emph{risk-based test process} and is aligned with the consideration of risks in all phases of the test process. In this section, we explain these classes, their subclasses and concrete values for each class of the risk-based testing taxonomy in depth. 

\subsection{Risk drivers}

As stated in~\cite{ISO14971}, risks result from hazards. Hazards in software-based systems stem from software vulnerabilities and from defects in software functionalities, which are critical to safety-related aspects or to the business cases of the software. One needs to test that a software-based system is 
\begin{itemize}
	\item reliable, i.e., able to deliver services as specified
	\item available, i.e., able to deliver services when requested
	\item safe, i.e., able to operate without harmful states
	\item secure, i.e., able to remain protected against accidental or deliberate attacks
	\item resilient, i.e., able to recover timely from unexpected events
	\end{itemize}
Since the respective test methods, i.e., functional testing, security testing, and performance testing differ, the considered risk drivers determine which testing is appropriate and has to be chosen. Therefore, the taxonomy begins with the \emph{risk drivers} to be a first differentiating element of risk-based testing approaches. We consider \emph{functionality}, \emph{security}, and \emph{safety} to be the dominant risk drivers for software. They together form the reliability, availability, safety, security, and resilience of a software-based system and hence constitute the options for the risk drivers in the RBT taxonomy. 

\subsection{Risk assessment}

The second differentiating element of RBT approaches is the way risks are being determined. According to~\cite{istqb2012standardGlossary}, \emph{risk assessment} is the process of \emph{identifying} and subsequently \emph{analyzing} the identified risk to determine its level of risk, typically by assigning likelihood and impact ratings. Risk assessment itself has multiple aspects, so that one needs to differentiate further into the \emph{risk item type}, to which the risk relate to, the \emph{factors} influencing risks, the \emph{risk estimation} technique used to estimate and/or evaluate the risk, and the \emph{degree of automation} for risk assessment.

\subsubsection{Risk item type} 

The risk item type determines the elements to which risk exposures and tests are assigned~\cite{felderer2013experiences}. Risk items can be of type \emph{generic risk}, i.e., risk items independent of a particular artifact like security risks or specific faults, \emph{test case}~\cite{yoon2011test}, i.e., directly test cases themselves as in regression testing scenarios, \emph{runtime artifact} like deployed services, \emph{functional artifact} like requirements or features, \emph{architectural artifact} like component, or \emph{development artifact} like source code file. The risk item type is determined by the test level. For instance, functional or architectural artifacts are often used for system testing, and generic risks for security testing. 
In addition, we use the term \emph{artifact} to openly refer to other risk items used in requirements capturing, design, development, testing, deployment, and/or operation and maintenance, which all might relate to the identified risks.

\subsubsection{Factors} 

The risk factors quantify identified risks~\cite{bai2012risk}: \emph{Risk exposure} is the quantified potential for loss. It is calculated by the likelihood of risk occurrence multiplied by the potential loss, also called the impact. The risk exposure considers typically aspects like liability issues, property loss or damage, and product demand shifts. RBT approaches might also consider the specific aspect of \emph{likelihood} of occurrence, e.g., for test prioritization or selection or the specific aspect of \emph{impact rating} to determine test efforts needed to analyze the countermeasures in the software.

\subsubsection{Estimation}

The estimation technique determines how the risk exposure is actually estimated and can be \emph{expert judgment} or \emph{formal model}~\cite{Jorgensen2009EffortEstimation}. The essential difference between formal-model-based and expert-judgment-based effort estimation is the quantification step-that is, the final step that transforms the input into the risk estimate. Formal risk estimation models are based on a mechanical quantification step such as a formula or a test model. On the other hand, judgment-based estimation methods are based on a judgment-based quantification step-for example, what the expert believes is most risky. Judgment-based estimation processes range from pure “gut feelings” to structured, historical data including failure history and checklist-based estimation processes. 

In addition, any risk estimation uses a scale to determine the risk ``level''. This risk scale can be \emph{quantitative} or \emph{qualitative}. Quantitative risk values are numeric and allow computations, qualitative risk values can only be sorted and compared. An often used qualitatively scale for risk levels is low, medium, and high~\cite{wendland2012systematic}.

\subsubsection{Degree of automation} 

Risk assessment can be supported by automated methods and tools. For example, risk-oriented metrics can be measured \emph{manually} or \emph{automatically}. The manual measurement is often supported by strict guidelines and the automatic measurement is often performed via static analysis tools. Other examples for automated risk assessment include the derivation of risk exposures from formal risk models, see, for instance,~\cite{FredriksenKGSOD02}.

\subsection{Risk-based test process}

Based on the risks being determined and characterized, RBT follows the fundamental test process~\cite{istqb2012standardGlossary} or variations thereof. All activities and phases in a test process are impacted by the risk perspective taken in RBT. RBT specifics in test processes are outlined in the following subsections. 

\subsubsection{Risk-based test planning}

\emph{Test planning} establishes or updates the scope, approach, resources, and schedule of intended test activities. Amongst others, \emph{test objectives}, \emph{test prioritization and selection}, \emph{risk metrics}, and \emph{exit criteria}, which impact risk-based testing~\cite{redmill2005theory}, are determined.

\paragraph{Test objectives} ~\\

RBT requires focusing the testing activities and efforts based on the risk assessment of the particular product or of the project, in which it is developed. In simple words: if there is high risk, then there will be serious testing. If there is no risk, then there will be rather little testing. For example, products with high complexity, new technologies, many changes, many defects found earlier, developed by personnel with less experiences or lower qualification, or developed along new or renewed development processes may have a higher probability to fail and need to be tested more thoroughly. The reason to design or execute a test, i.e., a \emph{test objective}, can be related to the risk item to be tested, to the thread scenarios of a risk item or to the counter measures established to secure that risk item, see also Section~\ref{RBTdesign}. 

\paragraph{Test prioritization and selection} ~\\

In order to optimize the costs of testing and/or the quality and fault detection capability of testing, techniques for prioritizing, selecting, and minimizing tests as well as combinations thereof have been developed and are widely in use~\cite{Yoo:2012:RTM:2284811.2284813}. In the ranges of intolerable risk and ``As Low As Reasonably Practicable (ALARP)''\footnote{The ALARP principle is typically used for safety-critical, but also for mission-critical systems. It says that the residual risk shall be as low as reasonably practical.} risks, these techniques are used to identify tests for the risk-related test objectives determined before. For example, design-based approaches for test selection~\cite{Briand:2009:ART:1465742.1466092} and coverage-based approaches~\cite{Amland2000RBT} for test prioritization are well-suited for RBT. 

\paragraph{Risk metrics} ~\\

Metrics are used to quantify different aspects in testing such as the minimum level of testing, extra testing needed because of high number of faults found, the quality of the tests and the test process. They are used to manage the RBT process and optimize it with respect to time, efforts, and quality~\cite{Amland2000RBT}.

\paragraph{Exit criteria}  ~\\

Typical exit criteria for testing that are used to report against and to plan when to stop testing, include all test ran successfully, all issues have been retested and signed off, or all acceptance criteria have been met. Specific RBT-related exit criteria~\cite{Amland2000RBT} add criteria on the residual risk in the product and coverage-related criteria: all risk items, their threat scenarios and/or counter measures being covered.

\subsubsection{Risk-based test design}
\label{RBTdesign}

\emph{Test design} is the process of transforming test objectives into test cases. This transformation is guided by the coverage criteria, which are used to quantitatively characterize the test cases and often used for exit criteria. Furthermore, the technique of transformation depend on the test types needed to realize a test objective. RBT uses coverage criteria specific to the risk artifacts and test types specific to the risk drivers on functionality, security, and safety.

\paragraph{Coverage criteria}  ~\\

In RBT, the classical code-oriented and model-based coverage criteria like path coverage, condition-oriented coverage criteria like modified condition decision coverage, requirements-oriented coverage criteria like requirements or use case coverage are extended with coverage criteria to cover selected or all assets, thread scenarios, and counter measures~\cite{Stallbaum:2008:ATR:1370042.1370057}. While asset coverage rather belongs to requirements-oriented coverage~\cite{wendland2012systematic}, thread scenario, and counter measure coverage can be addressed by code-oriented, model-based, and/or condition-oriented coverage criteria~\cite{hosseingholizadeh2010source}.

\paragraph{Test types}  ~\\

As reported by different computer emergency response teams such as GovCERT-UK, software defects continue to be a major, if not the main source of incidents caused by software-based systems. Therefore, functional testing is likewise a major test type in RBT to analyze reliability and safety aspects, see, e.g.,~\cite{Amland2000RBT}. In addition, security testing including penetration testing, fuzz testing and/or randomized testing is key in RBT~\cite{zech2011risk} to analyze security and resilience aspects. Furthermore, performance and scalability testing focusing on normal load, maximal load, and overload scenarios to analyze availability and resilience aspects, see, e.g.,~\cite{Amland2000RBT}.

\subsubsection{Risk-based test implementation}

\emph{Test implementation} comprises tasks like preparing test harnesses and test data, providing logging support or writing automated test scripts to enable the automated execution of test cases~\cite{istqb2012standardGlossary}. Risk aspects are especially essential for providing \emph{logging support} and for \emph{test automation}. 

\paragraph{Logging support} ~\\

Logging is the process of recording information about tests executed into a test log. Especially for risk-based testing, it is important to document the test progress via test logging~\cite{Amland2000RBT}. This may require adaptations of the logging support to meet the special requirements of risk-based testing, for instance, on risk items.

\paragraph{Test automation} ~\\

Test automation is the use of special software (separate from the software under test) to control the execution of tests and the comparison of actual outcomes with predicted outcomes~\cite{huizinga2007automated}. Experiences from test automation~\cite{graham2012experiences} show possible benefits like improved regression testing or a positive return on investment, but also caveats like high initial investments or difficulties in test maintenance. Risks may therefore be beneficial to guide decisions where and to what degree testing should be automated.   

\subsubsection{Risk-based test execution}

\emph{Test execution} is the process of running test cases. In this phase, risk-based testing is supported by \emph{monitoring} and \emph{risk metrics measurement}.

\paragraph{Monitoring} ~\\

Monitoring is run concurrently with a system under test and supervises, records or analyzes the behavior of the running system~\cite{Radatz1990IEEEStandardGlossary,istqb2012standardGlossary}. Differing from software testing, which actively stimulates the system under test, monitoring only passively observes a running system. For risk-based testing purposes, monitoring enables additional complex analysis, e.g., of the internal state of a system for security testing, as well as tracking the project's progress toward resolving its risks and taking corrective action where appropriate.

\paragraph{Risk metrics measurement} ~\\

\emph{Risk metrics measurement} determines risk metrics defined in the test planning phase. A measured risk metrics could be the number of observed critical failures for risk items where failure has high impact~\cite{felderer2013using}.

\subsubsection{Risk-based test evaluation}

\emph{Test evaluation} comprises decisions on the basis of exit criteria and logged test results compiled in a test report. In this respect, risks are \emph{mitigated} and may require a \emph{re-assessment}. Furthermore, risks may guide \emph{test exit decisions} and \emph{reporting}.

\paragraph{Risk reporting} ~\\

Test reports are documents summarizing testing activities and results~\cite{istqb2012standardGlossary} that communicate risks and alternatives requiring a decision. They typically report progress of testing activities against a baseline (such as the original test plan) or test results against exit criteria. In \emph{risk reporting}, assessed risks which are monitored during the test process, are explicitly reported in relation to other test artifacts. Risk reports can be descriptive summarizing relationships of the data or predictive using data and analytical techniques to determine the probable future risk. Typical descriptive risk reporting techniques are risk burn down charts which visualize the development of the overall risk per iteration as well as traffic light reports providing a high level view on risks using colors red for high risks, yellow for medium risks and green for low risks. A typical predictive risk reporting technique is residual risk estimation, for instance, based on software reliability growth models~\cite{goel1985reliability-models}.

\paragraph{Risk re-assessment} ~\\

The \emph{re-assessment of risks} after test execution may be planned in the process or triggered by a comparison of test results against the assessed risks. This may reveal deviations between the assessed and the actual risk level and require a re-assessment to adjust them. Test results can explicitly be integrated into a formal risk analysis model~\cite{stallbaum2007employing} or just trigger the re-assessment in an informal way.

\paragraph{Test exit decision} ~\\

The \emph{test exit decision} determines if and when to stop testing~\cite{felderer2013experiences}, but may also trigger further risk mitigation measures. This decision may be taken on the basis of a test report matching test results and exit criteria or ad hoc, for instance, solely on the basis of the observed test results.

\paragraph{Risk mitigation} ~\\

\emph{Risk mitigation} covers efforts taken to reduce either the likelihood or impact of a risk~\cite{571734}. In the context of risk-based testing, the assessed risks and their relationship to test results and exit criteria (which may be outlined in the test report), may trigger additional measures to reduce either the likelihood or impact of a risk to occur in the field. Such measures may be bug fixing, re-design of test cases or re-execution of test cases.

\section{Articles of this special section} \label{sec:sttt}

This special section on risk-based testing comprises four articles. Two of them are primary studies on risk-based testing~\cite{Neubauer2014ActiveContinuousQualityControl,Carrozza2014DynamicTestPlanning} presenting novel risk-based testing approaches, and two are secondary studies on risk-based testing~\cite{Felderer2014MultipleCaseStudy,Erdogan2014SLR} empirically evaluating existing risk-based testing approaches. In this section, we provide an overview of each article and discuss each of them in context of the proposed risk-based testing taxonomy. For this purpose, primary studies on risk-based testing are classified according to the taxonomy and secondary studies are used to additionally check its adequacy. \newline

The article ``Risk-based testing via active continuous quality control'' by Neubauer et al.~\cite{Neubauer2014ActiveContinuousQualityControl} shows how Active Continuous Quality Control (ACQC)~\cite{windmuller2013ACQC}, which employs learning technology to automatically maintain test models during system evolution, can be extended by risk assessment to support risk-based (regression) testing. Key to this enhancement is the tailoring of ACQC's characteristic model extraction based on automata learning to prioritize critical aspects. Technically, so called risk analysts are provided with an abstract modeling level tailored to design the required learning alphabets that encompass data flow constraints reflecting risk profiles at the user level. The resulting alphabet models steer the ACQC process in a fashion that increases risk coverage while it at the same time reduces the regression testing effort. This is illustrated by the application of the approach to Springer's Online Conference Service in two system evolution scenarios, i.e., system migration and functional extension. 

According to the risk-based testing taxonomy, the risk driver of the presented approach to risk-based testing via ACQC is fault-prone functionality expressed by symbols which reflect single critical actions of the SUT from the user perspective and together form an alphabet. Risk assessment itself is not directly addressed in the article. But it is stated that, ``a risk analyst uses his knowledge to select, parametrize (instantiate), and combine generic alphabet symbols to build tailored risk-based alphabet models'', and further, that, these alphabet models ``encompass data flow constraints reflecting risk profiles at the user level''. From these statements, it can be concluded that risks are assigned to functional artifacts, i.e., user actions, probability and impact factors are not defined, and risk exposure is measured automatically on a qualitative scale based on expert judgment. The risk-based alphabet models steer the ACQC process and are considered for test planning and design. The approach determines the generated test model which influences test selection and prioritization in the test planning phase, as well as coverage of risk items, i.e., critical actions. \newline

The article ``Dynamic test planning: a study in an industrial context'' by Carrozza et al.~\cite{Carrozza2014DynamicTestPlanning} presents a method to dynamically allocate testing resources to software components minimizing residual risks in terms of the estimated number of defects or estimated residual defect density. The method is based on software reliability growth models~\cite{goel1985reliability-models}, used at component level to monitor the testing progress of each component. From these models, an estimate of the quality achievable for a component in relation to the testing effort devoted to it is obtained. By iteratively solving an optimization problem, the testing effort is consecutively directed towards the component contributing the most to reduce the number of residual defects or their density in the overall system. The method is implemented in a tool and applied to a real-world critical system in the homeland security domain aiming at the management of port, maritime and coastal surveillance. Results of the application show improved test process outcome measured in terms of detected defects compared to uniform or size-based allocation of testing resource given a predefined amount of testing resources. 

With regard to the taxonomy, the presentation of the method in the article focuses on risk assessment to dynamically allocate testing resources, but its integration into the test process is only addressed on the side. The considered risk driver is again fault-prone functionality. Risk is assessed for architectural artifacts, i.e., software components, taking probability based on defects into account. Estimation is based on a formal analysis model using software reliability growth models to quantitatively measure risks on the basis of the estimated number of residual defects and their density. The measurement is automated by a tool presented in the article. The assessed risk of components is applied in the test planning phase as a metrics to distribute testing resources to components. During test execution, risk in terms of observed defects is monitored and taken into account to re-assess and report risks in the test evaluation phase. \newline

The article ``A multiple case study on risk-based testing in industry'' by Felderer and Ramler~\cite{Felderer2014MultipleCaseStudy} presents a case study on currently applied risk-based testing approaches grounded on three industry cases from different backgrounds, i.e., a test project in context of the extension of a large information system, product testing of a measurement and diagnostic equipment for the electrical power industry as well as a test process of a system integrator of telecommunications solutions. The main analysis across the three cases was conducted qualitatively by exploring documents, tools and interview protocols. The study revealed that all cases follow a common definition of risk relating its likelihood and impact. However, the definition may remain implicit and the degree of formality depends on the application scope, i.e., it increases from project to product, and, furthermore, to process. In the three cases, risk is taken into consideration in virtually all testing activities. Risks are calculated from risk information associated to testable entities for identifying critical areas of the software system to focus testing effort on. Risk-based testing relies on the established standard tool infrastructure for testing which is partially supported by tools for collecting and analyzing measurement data. Finally, as central benefits of risk-based testing using risk information to increase the range of testing for detecting additional defects, and to target the most critical defects from the very beginning, as well as the incorporation of risks for informed decision-making in testing are identified. 

The findings from the multiple case study on risk-based testing substantiate the defined taxonomy. The observation that risks are taken into account in almost all phases of the test process justifies the alignment of the taxonomy with the consideration of risks in the test process phases. In addition, the categories risk item type, factors, estimation and degree of automation, defined for the class risk assessment as well as their values, are confirmed by the findings following findings already mentioned before: (1) all cases follow a common definition of risk relating its likelihood and impact, (2) risks are calculated from risk information associated to testable entities for identifying critical areas to focus testing effort on, (3) the degree of formality of the risk definition depends on the application scope, and (4) risk-based testing is partially supported by tools for collecting and analyzing measurement data. \newline

The article ``Approaches for the combined use of risk analysis and testing: a systematic literature review'' by Erdogan et al.~\cite{Erdogan2014SLR} provides a systematic literature review on risk-based testing approaches, which use risk analysis to support testing, as well as test-based risk analysis approaches, which use testing to support risk analysis. In the article, the term ``risk analysis'' is used in the sense of risk assessment. The authors follow the guidelines of Kitchenham and Charters~\cite{kitchenham2007SLR} for conducting a systematic literature review. From the search engines Google Scholar, IEEE Xplore Digital Library, SpringerLink, ScienceDirect, and ACM Digital Library a total of 32 peer-reviewed papers reporting approaches for the combined use of risk assessment and testing were selected for inclusion in the survey and grouped by the first author into 24 approaches which were considered for further analysis with the following results: Only two approaches address test-based risk assessment, while the remaining 22 focus solely on various aspects of risk-based testing. From the 22 risk-based testing approaches, four address the combination of risk assessment and testing at a general level, two address model-based risk estimation, five focus on test case generation, three focus on test case analysis, one addresses source code analysis, and one aims at measurement. Besides these six generic types of RBT approaches, specific RBT approaches were identified, i.e., two addressing specific programming paradigms and four targeting specific applications. The analysis further shows that recurring goals of the RBT approaches are to improve effectiveness of testing and to reduce time and costs. Safety and security are in particular of special concern in test case generation approaches. Most approaches are general and not intended to be used in a specific context. For the approaches targeting specific applications, web services/applications and cloud computing are the most common types. The level of formality varies from purely qualitative risk levels without initial risk identification and analysis to rigorous mathematical models. Tool support is rarely reported as only one approach presents a complete tool for automatically generating test cases and four approaches are supported by tool prototypes.  

The systematic literature review provides evidence for the values of the risk assessment categories estimation and degree of automation in the RBT taxonomy. Furthermore, each of the generic types of RBT approaches identified in the systematic literature review can be aligned with the taxonomy: 
\begin{itemize}
  \item Approaches addressing the combination of risk assessment and testing at a general level directly follow the top-level classification principle of the taxonomy combining risk assessment with the integration of risk into the test process.
  \item Approaches with main focus on model-based risk estimation address a formal risk analysis model.
  \item Approaches with main focus on test case generation address risk-based test design and implementation.
  \item Approaches with main focus on test case analysis focus on selection and prioritization of generated test cases.
  \item Approaches based on automatic source code analysis address automatic risk assessment of the factor probability for development artifacts by quantitative and formal estimation.  
  \item Approaches aiming at measurement address the classes risk assessment and risk-based test evaluation in the taxonomy.
\end{itemize}

Finally, all specific RBT approaches take risk assessment for specific programming paradigms or applications into account, and, besides that, mainly address test selection and prioritization~\cite{kumar2009enabling,bai2009risk,murthy2009leveraging} or test design~\cite{casado2010testing,zech2011risk}. Only Rosenberg~\cite{rosenberg2000risk} addresses solely risk assessment (aiming at risk-based testing) for a specific programming paradigm, i.e., object-oriented systems.

\section{Conclusion} \label{sec:conclusion}

In this paper, we presented a taxonomy of risk-based testing. It is aligned with the consideration of risks in all phases of the test process and consists of three top-level classes, i.e., risk drivers, risk assessment, and risk-based test process. Risk drivers can be functionality, safety, and security. Risk assessment comprises the subclasses risk item type, factors, estimation, and degree of automation. The risk-based test process then takes the assessed risks into account to guide the activities of the test process providing risk-based test planning, design, implementation, execution, and evaluation. The taxonomy provides a framework to understand, categorize, assess and compare risk-based testing approaches to support their selection and tailoring for specific purposes. We further presented the four articles of the STTT Special Section on Risk-Based Testing and discussed them in the context of the taxonomy. Two articles of the special section~\cite{Neubauer2014ActiveContinuousQualityControl,Carrozza2014DynamicTestPlanning} present novel risk-based testing approaches and were classified according to the taxonomy. The two other articles~\cite{Felderer2014MultipleCaseStudy,Erdogan2014SLR} comprise secondary studies on risk-based testing. These studies empirically evaluate existing risk-based testing approaches and were used to additionally check the presented taxonomy of risk-based testing.  
  
\section*{Acknowledgments}

This research was partially funded by the research projects MOBSTECO (FWF P 26194-N15), QE LaB - Living Models for Open Systems (FFG 822740), ITEA2 DIAMONDS (Development and Industrial Application of Multi-Domain-Security Testing Technologies), and EU RASEN (Compositional Risk Assessment and Security Testing of Networked Systems).

\balance
\bibliographystyle{splncs03}
\bibliography{references} 

\end{document}